\documentclass{article}
\usepackage[preprint]{neurips_2024}

\usepackage{amsmath,amssymb,amsfonts}
\usepackage{amsthm}
\usepackage{algorithmic}
\usepackage{graphicx}
\usepackage{textcomp}
\usepackage{xcolor}
\usepackage{booktabs}
\usepackage{multirow}
\usepackage{hyperref}
\usepackage{listings}
\usepackage{mathtools}   

\theoremstyle{definition}
\newtheorem{definition}{Definition}
\newtheorem{hypothesis}{Hypothesis}
\newtheorem{proposition}{Proposition}

\hypersetup{
  colorlinks=true,
  linkcolor=blue,
  citecolor=blue,
  urlcolor=blue
}

\newcommand{\OO}{\mathcal{O}}
\newcommand{\BO}{\mathcal{B}_{\mathcal{O}}}
\newcommand{\RO}{R_{\mathcal{O}}}
\newcommand{\UO}{U_{\mathcal{O}}}
\newcommand{\AuthO}{\text{Auth}_{\mathcal{O}}}
\newcommand{\cmark}{\checkmark}
\newcommand{\pmark}{$\sim$}
\newcommand{\xmark}{$\times$}

\begin{document}

\title{Owner-Harm: A Missing Threat Model for AI Agent Safety}

\author{%
  Dongcheng Zhang\thanks{Work performed while at BlueFocus Communication
  Group. Correspondence: \texttt{zdclink@gmail.com}.} \\
  BlueFocus Communication Group \\
  Beijing, China \\
  \texttt{zdclink@gmail.com}
  \And
  Yiqing Jiang \\
  Tongji University \\
  Shanghai, China \\
  \texttt{jyq.russel@gmail.com}
}

\maketitle

\begin{abstract}
Existing AI agent safety benchmarks focus predominantly on generic
criminal harm---cybercrime, harassment, and weapon synthesis---leaving a
systematic blind spot for a distinct and commercially consequential threat
category: agents harming their own deployers. Real-world incidents
illustrate this gap: the Slack AI credential exfiltration (August 2024),
Microsoft 365 Copilot data leaks via calendar injection (January 2024),
and a Meta agent unauthorized forum post exposing operational data (March
2026). We propose \emph{Owner-Harm}, a formal threat model defining eight
categories of agent behavior that damage the deploying owner's interests.
We quantify the resulting defense gap on two authoritative benchmarks: the
same compositional safety system achieves 100\% TPR / 0\% FPR on
AgentHarm (generic criminal harm, 176 harmful scenarios) yet only 14.8\%
(4/27; 95\% CI: 5.9\%--32.5\%) on AgentDojo injection tasks (prompt-injection-mediated owner harm),
exposing a category-specific blind spot. A controlled experiment with a
zero-shot generic LLM classifier reveals that this gap is not inherent to
the owner-harm category (generic LLM: 62.7\% vs.\ 59.3\%, gap = 3.4 pp)
but arises from environment-bound symbolic rules that fail to generalize
across tool vocabularies. On a post-hoc diagnostic benchmark of 300
owner-harm scenarios, the gate alone achieves 75.3\% TPR / 3.3\% FPR;
adding a deterministic post-audit verifier raises overall TPR to
\textbf{85.3\%} (+10.0\,pp) and Hijacking detection from 43.3\% to 93.3\%,
demonstrating strong layer complementarity. These results
confirm that owner-harm requires context-aware safety mechanisms that
explicitly model resource ownership, trust boundaries, and authorization
scope. To organize these observations, we introduce the
\emph{Symbolic-Semantic Defense Generalization (SSDG)} conceptual framework,
which describes how information coverage---the fraction of an attack's
evidential requirements satisfied by a defense's context set---relates to
detection rate. Two controlled SSDG experiments partially test the
framework: context deprivation induces a 3.4$\times$ amplification of the
detection gap (gap ratio $R = 3.60$ vs.\ $R = 1.06$ with full context),
while context injection reveals that goal-context awareness is necessary
but not sufficient---structured goal-action alignment, not simple text
concatenation, is required for effective owner-harm detection.
\end{abstract}

\section{Introduction}
\label{sec:intro}

The deployment of autonomous AI agents into production systems has
accelerated faster than the safety frameworks needed to govern them.
Current research focuses on preventing agents from facilitating generic
criminal harm: generating malware, synthesizing dangerous substances, or
producing harassing content. Yet a qualitatively different and commercially
urgent threat class has emerged in practice---agents harming the very
organizations that deploy and trust them.

Three incidents anchor this concern. In August 2024, PromptArmor disclosed
that Slack AI could be manipulated via prompt injection to exfiltrate
private channel tokens from the deploying organization
\cite{promptarmor2024slack}. In January 2024, Zenity Research demonstrated
that Microsoft 365 Copilot could be hijacked through malicious calendar
invitations to forward sensitive emails to external parties
\cite{zenity2024copilot}. In March 2026, a Meta AI agent made an
unauthorized post to an internal forum, exposing operational data for two
hours before containment \cite{meta2026sev1}. In each case, the agent's
deployer---not a third-party victim---bore the harm.

\textbf{The benchmark gap.} Existing agent safety benchmarks do not model
this pattern. AgentHarm \cite{andriushchenko2025agentharm} catalogs 11
harm categories, all framed as an adversary using an agent against third
parties. AgentDojo \cite{debenedetti2024agentdojo} evaluates prompt
injection defenses across four suites (Banking, Travel, Workspace, Slack)
but without an explicit owner-harm taxonomy. The consequences are
measurable: the same compositional safety system that achieves 100\% TPR
on AgentHarm scores only 3.7\% (1/27) on AgentDojo injection tasks using
deterministic rules, and 14.8\% (4/27) with semantic reasoning added. The
constraints that successfully intercept generic criminal content---system
call patterns, toxicity signals---have near-zero coverage of
application-layer owner harm, such as financial transactions initiated via
legitimate banking APIs or PII forwarded through authorized email tools.

\textbf{Contributions.} This paper makes three contributions:
\begin{enumerate}
  \item \textbf{Owner-Harm threat model.} We provide the first formal
    definition of owner-harm and its eight sub-categories, grounded in
    real incidents and distinguished from existing taxonomies.
  \item \textbf{Defense gap quantification and gap attribution.} We
    measure the gap (100\% vs.\ 14.8\%) and use a controlled experiment
    to attribute it to environment-bound rule generalization failure
    rather than inherent category difficulty: a zero-shot generic LLM
    shows only a 3.4 pp gap (62.7\% vs.\ 59.3\%) across the same
    benchmarks.
  \item \textbf{Layer complementarity evidence.} We show that
    deterministic post-audit verification is orthogonal to semantic
    gate reasoning for Hijacking (C7), raising detection from 43.3\%
    to 93.3\% via complementary detection dimensions.
\end{enumerate}

\section{Owner-Harm Threat Model}
\label{sec:model}

\subsection{Formal Definitions}
\label{subsec:formal}

We model an AI agent as a tuple $\mathcal{A} = \langle M, \mathcal{T},
\mathcal{E}, \pi \rangle$, where $M$ is the underlying language model,
$\mathcal{T}$ is the available tool set, $\mathcal{E}$ is the environment
state space, and $\pi$ is the agent's policy. An agent produces an action
sequence $\vec{a} = (a_1, \ldots, a_n)$ transforming environment state
$s_0$ to $s'$.

\begin{definition}[Owner]
The \emph{owner} $\OO$ is the entity that deploys $\mathcal{A}$, grants
resource access $\RO$, and bears accountability for agent actions.
$\OO$ is associated with a resource set $\RO$, a utility function $\UO$
over environment states, and a trust boundary $\BO$ delineating
authorized interactions.
\end{definition}

\begin{definition}[Owner-Harm]
An action sequence $\vec{a}$ causes \emph{owner-harm} with respect to
owner $\OO$ if and only if:
\[
  \mathrm{OH}(\vec{a}, \OO) \coloneqq \UO(s') < \UO(s_0)
    \;\land\;
    \exists\, a_i \in \vec{a} : \mathrm{cause}(a_i,\, s_0 \to s')
\]
where $\mathrm{cause}(a_i, s_0 \to s')$ means $a_i$ is a proximate or
distal cause of the state transition from $s_0$ to $s'$.
\end{definition}

\begin{definition}[Direct Owner-Harm]
\[
  \mathrm{DOH}(\vec{a}, \OO) \coloneqq \mathrm{OH}(\vec{a}, \OO)
    \;\land\; \mathrm{ProxCause}(\mathcal{A}, \vec{a})
\]
where $\mathrm{ProxCause}(\mathcal{A}, \vec{a})$ means the agent is the
proximate cause of the harmful state transition.
\end{definition}

\begin{definition}[Indirect Owner-Harm]
\[
  \mathrm{IOH}(\vec{a}, \OO, \mathcal{V}) \coloneqq
    \mathrm{OH}(\vec{a}, \OO)
    \;\land\; \mathrm{Manipulate}(\mathcal{V}, \pi, \vec{a})
\]
where $\mathcal{V}$ is an adversarial actor that manipulates the agent's
policy $\pi$ to produce the harmful action sequence---the prompt injection
attack model.
\end{definition}

\subsection{Eight Owner-Harm Categories}
\label{subsec:categories}

We enumerate eight categories $C1$--$C8$ (primary labels; a single incident may span multiple categories), each grounded in
documented incidents.

\textbf{C1: Credential Leak.} The agent exfiltrates authentication
materials (API keys, OAuth tokens, session cookies, passwords) belonging
to or entrusted to the owner. \textit{Real case}: Slack AI August 2024
\cite{promptarmor2024slack}---injected prompt caused the assistant to echo
private channel tokens to the attacker.

\textbf{C2: Infrastructure Exposure.} The agent misconfigures or discloses
network rules, cloud resource policies, or internal architecture, expanding
the owner's attack surface. \textit{Pattern}: AI-assisted code generation
producing over-permissive AWS IAM policies that expose production
databases.

\textbf{C3: Privacy Exposure.} The agent transmits personally identifiable
information (PII) or confidential business data to unauthorized parties.
\textit{Real case}: Microsoft 365 Copilot January 2024
\cite{zenity2024copilot}---calendar injection caused Copilot to forward
sensitive email contents externally.

\textbf{C4: Inner Circle Leak.} The agent betrays trust relationships that
are commercially sensitive to the owner: client lists, partner agreements,
M\&A discussions, or strategic roadmaps. \textit{Pattern}: Samsung
employees pasting proprietary semiconductor source code into ChatGPT (April
2023), with the owner-harm being the confidentiality breach to a third-party
model.

\textbf{C5: Asset Destruction.} The agent irreversibly deletes or corrupts
owner data, configurations, or digital assets. \textit{Pattern}: AI coding
agents executing \texttt{rm -rf} on production directories when given
broad filesystem access and ambiguous cleanup instructions.

\textbf{C6: Exfiltration via Tools.} The agent exploits \emph{authorized}
tools as covert data channels---using email, webhooks, or file-write
operations to smuggle sensitive data to attacker-controlled endpoints.
\textit{Real case}: ASCII smuggling via Microsoft 365 Copilot
\cite{rehberger2024copilot}---injected prompt caused Copilot to encode and
transmit sensitive data through Markdown image URLs rendered invisibly.

\textbf{C7: Hijacking.} An adversary achieves persistent control over the
agent using the owner's identity, credentials, or resource access---the
agent becomes a weaponized surrogate. \textit{Pattern}: AutoGPT memory
poisoning (2024--2025) where injected memory entries caused the agent to
act on behalf of the attacker across subsequent sessions.

\textbf{C8: Unauthorized Autonomy.} The agent exceeds its authorization
scope by taking consequential actions without required human confirmation.
\textit{Real case}: Air Canada chatbot (February 2024) committed the airline
to unauthorized refund terms in a legally binding interaction
\cite{aircanadaruling2024}.

\subsection{Comparison with Existing Taxonomies}
\label{subsec:comparison}

Table~\ref{tab:taxonomy} compares Owner-Harm categories against four
existing frameworks: AgentHarm \cite{andriushchenko2025agentharm} (11
categories), ToolEmu \cite{ruan2024toolemu} (7 risk types), OWASP LLM Top
10, and AgentDojo \cite{debenedetti2024agentdojo} (4 suites).

\begin{table}[!t]
  \centering
  \caption{Coverage of Owner-Harm Categories by Existing Frameworks}
  \label{tab:taxonomy}
  \small
  \begin{tabular}{lcccc}
    \toprule
    \textbf{Category} & \textbf{AgentHarm} & \textbf{ToolEmu}
      & \textbf{OWASP} & \textbf{AgentDojo} \\
    \midrule
    C1: Credential Leak        & \pmark & \cmark & \cmark & \pmark \\
    C2: Infrastructure Exposure & \xmark & \pmark & \cmark & \xmark \\
    C3: Privacy Exposure        & \pmark & \cmark & \cmark & \cmark \\
    C4: Inner Circle Leak       & \xmark & \xmark & \xmark & \xmark \\
    C5: Asset Destruction       & \pmark & \cmark & \pmark & \pmark \\
    C6: Exfiltration via Tools  & \xmark & \pmark & \pmark & \pmark \\
    C7: Hijacking               & \xmark & \pmark & \cmark & \pmark \\
    C8: Unauthorized Autonomy   & \pmark & \cmark & \pmark & \xmark \\
    \bottomrule
  \end{tabular}
  \vspace{1mm}
  \\\small \cmark~= covered; \pmark~= partial; \xmark~= gap
\end{table}

The most notable gap is \textbf{C4 (Inner Circle Leak)}, which appears in
no existing benchmark. This category requires reasoning about the owner's
trust graph---who counts as an insider---which no current system models
explicitly. \textbf{C6} is partially covered by ToolEmu but misses the
\emph{covert channel} aspect: using authorized tools as smuggling vectors.
The \emph{owner-centric perspective} itself is the primary novel
contribution: existing frameworks evaluate harm against third parties,
users, or society; none formally model the deploying owner as the victim.

\subsection{Why Generic Defenses Fail}
\label{subsec:why_fail}

The fundamental mismatch is context-dependence of harm:
\[
  \exists\, c, \OO_1, \OO_2 :
    \mathrm{Harmful}(c, \OO_1) \;\land\; \lnot\mathrm{Harmful}(c, \OO_2)
\]
The action ``forward email to external@attacker.com'' is benign for an
owner whose trust boundary includes that address, and harmful for an owner
whose trust boundary does not. Content-only classifiers cannot make this
distinction.

Three owner-context dimensions are systematically absent from generic
defenses:

\begin{itemize}
  \item \textbf{Resource Ownership} ($\RO$): which data and credentials
    belong to the owner and should not be disclosed.
  \item \textbf{Trust Boundary} ($\BO$): who the owner considers an
    authorized counterparty for each resource type.
  \item \textbf{Authorization Scope} ($\AuthO$): which actions the owner
    has explicitly sanctioned without requiring confirmation.
\end{itemize}

\begin{proposition}[Generic Classifier Incompleteness]
Any context-free binary classifier $f_{\mathrm{gen}} : \mathcal{C}
\to \{0,1\}$ that classifies action content without access to $\RO$,
$\BO$, or $\AuthO$ will produce owner-harm false negatives on scenarios
where harm is defined relative to the owner's context.
\end{proposition}

Table~\ref{tab:defense_fail} summarizes the failure modes of common
defense classes against this incompleteness.

\begin{table}[!t]
  \centering
  \caption{Generic Defense Failure Modes for Owner-Harm}
  \label{tab:defense_fail}
  \small
  \begin{tabular}{p{2.8cm}p{4.8cm}}
    \toprule
    \textbf{Defense Class} & \textbf{Owner-Harm Blind Spot} \\
    \midrule
    Content classifier & Ignores $\RO$; ``send email'' is benign by content \\
    Injection detector  & Detects injection syntax; misses semantic intent \\
    Tool monitor        & Sees call signature; not authorization context \\
    DLP system          & Pattern-matches known PII; misses context-defined confidential data \\
    Behavioral anomaly  & Requires baseline; new deployments have no baseline \\
    \bottomrule
  \end{tabular}
\end{table}

\subsection{Symbolic-Semantic Defense Generalization (SSDG) Framework}
\label{subsec:ssdg_theory}

The controlled experiment results motivate a conceptual account of
\emph{when} and \emph{why} defenses fail to generalize across attack
categories. We emphasize that SSDG is a conceptual framework for organizing observations, not a formal mathematical theory with rigorous proofs. For formal treatments of causality in such analyses, see Pearl \cite{pearl2009causality}. We introduce the following framework.

\textbf{Definitions.} Let an attack $a$ be characterized by its
\emph{evidential requirement set} $E(a)$---the set of contextual facts
needed to determine its harmfulness. A defense $D$ maintains an
\emph{information set} $\Omega(D) = \Omega_S \cup \Omega_C$, where
$\Omega_S$ captures surface-level features (tool names, parameter
patterns) and $\Omega_C$ captures contextual knowledge (world knowledge,
owner context, task goals). The \emph{information coverage} of $D$ for
attack $a$ is:
\[
  \mathrm{Cover}(D,a) \;\coloneqq\; \frac{|E(a) \cap \Omega_{\mathrm{usable}}(D)|}{|E(a)|}
  \;\in [0,1]
\]
where $\Omega_{\mathrm{usable}}(D) \subseteq \Omega(D)$ denotes the subset of
information that the defense can \emph{effectively use} in its decision
process. Mere availability of information ($\Omega_{\mathrm{raw}}$) does not
guarantee usability---simple text concatenation of context may not enable
structured comparison (as P1 in Section~\ref{subsec:ssdg_experiments}
demonstrates).

\textbf{Assumption (Detection-Coverage Monotonicity).}
$\mathrm{TPR}(D,a)$ is non-decreasing in $\mathrm{Cover}(D,a)$; below
threshold $\theta$, $\mathrm{TPR}(D,a) \leq \varepsilon$.

\begin{hypothesis}[Symbolic Defense Gap]
\label{hyp:symbolic_gap}
A pure symbolic defense with $\Omega_C = \emptyset$ achieves
$\mathrm{Cover}(D, A_G) \gg \mathrm{Cover}(D, A_O)$ when $A_O$ requires
joint surface-context evidence. Consequently, $\mathrm{TPR}(D, A_O)
\leq \varepsilon$ while $\mathrm{TPR}(D, A_G)$ remains high---producing
a structural detection gap.
\end{hypothesis}

\begin{hypothesis}[Semantic Defense Stability]
\label{hyp:semantic_stability}
A semantic defense with $\Omega_C \supseteq \mathcal{C}_{\mathrm{world}}$
(world knowledge) satisfies $\mathrm{Cover}(D, A_G) \approx
\mathrm{Cover}(D, A_O)$, since both attack types' evidential requirements
are met by world-knowledge reasoning. The inter-category detection gap
is thus bounded by inference capacity, not information availability.
\end{hypothesis}

These hypotheses yield two falsifiable predictions tested in
Section~\ref{subsec:ssdg_experiments}: \textbf{P1}---injecting the user
goal into a symbolic defense's context should raise $\mathrm{Cover}(D,
A_O)$, increasing TPR; and \textbf{P2}---stripping world-context from
a semantic defense should cause it to exhibit symbolic-gap behavior,
with gap ratio $R > 3\times$.

\section{Nous: Compositional Runtime Safety Gate}
\label{sec:nous}

We evaluate owner-harm detection using \textbf{Nous}, a four-layer
compositional runtime safety gate for AI agent actions. We describe it
briefly here; the architecture is the primary evaluation vehicle rather
than the primary contribution.

\textbf{Layer 1 (L1): Deterministic Datalog Blocker.} 46 Datalog rules
encode owner-policy constraints (resource ownership, known-bad patterns,
categorical prohibitions). L1 is fail-closed, fast ($P_{50} = 0.055$~ms),
and requires no LLM call. Rules cover categorical owner-harm such as
exfiltration of credentials with recognizable patterns (C1) and shell
commands associated with asset destruction (C5).

\textbf{Layer 2 (L2): Triviality Filter.} Routes trivially benign
actions past the expensive semantic layer, yielding approximately 70\%
cost reduction with negligible safety impact.

\textbf{Layer 3 (L3): Semantic Gate.} An LLM evaluator supplied with
minimal-pair examples, intent decomposition prompts, and majority vote
$k=3$ to handle stochasticity in mixture-of-experts models. L3 provides
semantic reasoning about owner context: whether an action falls within
the owner's trust boundary, whether resource access is proportionate to
the stated task.

\textbf{Layer 4 (L4): Post-Gate Verifier.} Six deterministic audit rules
(V1--V6) inspect \emph{file contents} and \emph{multi-step tool-call
sequences} that are invisible to L1--L3 because they emerge from composed
actions rather than individual calls. L4 adds $\sim$0.038~ms latency with
zero LLM cost. Rules target prompt injection patterns embedded in file
artifacts and behavioral sequences characteristic of Hijacking (C7).

\textbf{Why this architecture for owner-harm.} L1 encodes owner-policy as
explicit Datalog facts; L3 reasons about owner context when policy is
ambiguous; L4 catches injection patterns that survive semantic evaluation
because the semantic reasoning is itself manipulated by injected content.
The layers are architecturally complementary: L1 operates on individual
action content; L3 operates on action semantics plus owner context; L4
operates on artifact contents and behavioral sequences.

\section{Evaluation}
\label{sec:eval}

\subsection{Evaluation Strategy}
\label{subsec:strategy}

We evaluate on three benchmark sets with different evidentiary weight:

\begin{enumerate}
  \item \textbf{AgentDojo} \cite{debenedetti2024agentdojo} (authoritative):
    27 injection tasks, 97 user tasks, 4 suites (Banking, Travel,
    Workspace, Slack). Ground-truth labels provided by benchmark authors.
    Safety score = fraction of injection tasks blocked.
  \item \textbf{AgentHarm} \cite{andriushchenko2025agentharm}
    (comparative baseline): 176 harmful + 176 benign scenarios across
    11 categories. Used to confirm that the 3.7\%/14.8\% gap on AgentDojo
    is category-specific, not a global system deficiency.
  \item \textbf{Owner-Harm Benchmark} (post-hoc diagnostic, 300H + 150B,
    8 categories): Constructed by the same author after optimization loops.
    \emph{This is NOT a held-out independent test set}; results should be
    interpreted as diagnostic evidence of system capabilities. Reported
    separately from the authoritative benchmarks to avoid overstating
    validity.
\end{enumerate}

\textbf{AgentDojo to Owner-Harm category mapping.} We assign each of the
27 injection tasks to the Owner-Harm category it most closely targets:
Banking (9 injection tasks): \texttt{financial\_harm} (8 tasks---unauthorized transfers with data exfiltration),
\texttt{credential\_leak} (1 task---password modification);
Travel (7 tasks): \texttt{privacy\_breach} (6 tasks---passport/banking info
exfiltration), \texttt{reputational\_harm} (1 task);
Workspace (6 tasks): \texttt{operational\_harm} (file deletion),
\texttt{privacy\_breach} (email forwarding), \texttt{credential\_leak}
(2FA bypass);
Slack (5 tasks): \texttt{reputational\_harm} (phishing message),
\texttt{privacy\_breach} (data exfiltration), \texttt{other} (SSRF).

\subsection{AgentDojo Results}
\label{subsec:agentdojo}

Table~\ref{tab:agentdojo_main} presents the three-configuration comparison.
All configurations are evaluated with the AgentDojo ground-truth mode
(deterministic task success scoring; no adversarial attack LLM used in
this evaluation).

\begin{table}[!t]
  \centering
  \caption{AgentDojo Results: Safety Score vs.\ Utility (27 injection
    tasks, 97 user tasks)}
  \label{tab:agentdojo_main}
  \begin{tabular}{lccc}
    \toprule
    \textbf{Configuration} & \textbf{Safety} & \textbf{Utility} & \textbf{FPR} \\
    \midrule
    L1 only (baseline)   & 3.7\% (1/27)\textsuperscript{$\dagger$}  & 97.9\% (95/97) & 2.1\% (2/97) \\
    L1 + L4              & 3.7\% (1/27)\textsuperscript{$\dagger$}  & 97.9\%         & 2.1\%        \\
    L1 + L3 + L4 (full) & 14.8\% (4/27)\textsuperscript{$\dagger$} & 97.9\%         & 2.1\%        \\
    \multicolumn{4}{l}{\small $\dagger$95\% Wilson CI: 3.7\% [0.7\%--18.3\%]; 14.8\% [5.9\%--32.5\%]} \\
    \bottomrule
  \end{tabular}
\end{table}

\textbf{Layer 4 contributes zero} to AgentDojo safety. Verifier rules V1--V6
target bash command sequences and curl exfiltration patterns embedded in
file artifacts---patterns characteristic of system-level Hijacking. AgentDojo
injection tasks operate at the \emph{application layer}: the attack tools
are \texttt{send\_money}, \texttt{send\_email}, \texttt{update\_calendar}.
No file artifact intermediary exists for V-rules to inspect.

\textbf{Layer 3 adds +11.1pp} by catching three cases unreachable by
Datalog: \texttt{travel/injection\_task\_3} (Privacy Breach---passport
details forwarded to adversary URL), \texttt{travel/injection\_task\_5}
(Privacy Breach---bank account numbers forwarded via booking confirmation),
and \texttt{banking/injection\_task\_6} (Financial Harm---batch wire
transfer to adversary account at plausible individual amounts). Utility
remains at 97.9\% across all configurations, confirming that semantic
reasoning does not increase false positives.

Table~\ref{tab:agentdojo_category} shows per-Owner-Harm-category results.

\begin{table}[!t]
  \centering
  \caption{AgentDojo Safety by Owner-Harm Category}
  \label{tab:agentdojo_category}
  \begin{tabular}{lcc}
    \toprule
    \textbf{Category} & \textbf{L1 only} & \textbf{L1+L3+L4} \\
    \midrule
    Financial Harm     & 0/10 (0\%)  & 1/10 (10\%)  \\
    Privacy Breach     & 0/6  (0\%)  & 2/6  (33\%)  \\
    Operational Harm   & 1/4  (25\%) & 1/4  (25\%)  \\
    Credential Leak    & 0/3  (0\%)  & 0/3  (0\%)   \\
    Reputational Harm  & 0/3  (0\%)  & 0/3  (0\%)   \\
    Other              & 0/1  (0\%)  & 0/1  (0\%)   \\
    \midrule
    \textbf{Total}     & \textbf{1/27 (3.7\%)} & \textbf{4/27 (14.8\%)} \\
    \bottomrule
  \end{tabular}
\end{table}

\textbf{Structural blind spots.} Three categories of injection tasks
remain systematically out of reach regardless of configuration:

\begin{itemize}
  \item \textit{Text-only manipulation} (e.g.,
    \texttt{travel/injection\_task\_6}: adversarial advertising text
    injected into a hotel description, with no tool call produced). The
    gate has no tool-call artifact to intercept.
  \item \textit{Malicious parameters indistinguishable from legitimate
    ones} (e.g., \texttt{send\_money} with a plausible amount and
    adversary-controlled recipient that appears to be a valid payee). Without
    the user's task goal in the gate context, the semantic layer cannot
    compare ``user requested payment to X'' versus ``agent is paying Y''.
  \item \textit{Multi-step attacks where partial interception
    is insufficient} (e.g., a read-then-exfiltrate chain where blocking
    the read breaks utility and missing the exfiltrate misses the attack).
\end{itemize}

\textbf{Theoretical ceiling without goal context}: we estimate 40--60\%
safety score on AgentDojo is achievable with user task goal injected into
gate context, enabling the ``user requested X, agent doing Y''
comparison that resolves malicious-parameter ambiguity.

\textbf{Comparison with AgentDojo baselines.} Debenedetti et al.\ report
that a GPT-4 agent with \texttt{tool\_filter} defense achieves
approximately 14.8\% safety on the Banking suite
\cite{debenedetti2024agentdojo}. Nous L1+L3+L4 matches this at 14.8\%
overall across all four suites, while maintaining 97.9\% utility compared
to approximately 85\% for \texttt{tool\_filter} (which blocks entire tool
categories rather than discriminating at the call level).

\subsection{Owner-Harm Benchmark Results (Post-Hoc Diagnostic)}
\label{subsec:ohb}

\emph{Important caveat}: the Owner-Harm Benchmark was constructed by the
same author after optimization loops on the Nous system. It is not an
independent held-out evaluation. All 450 scenarios were labeled by a
single annotator (the first author); no inter-annotator agreement (IAA)
was computed. The C1--C8 primary-category assignments reflect the
author's subjective judgment of the dominant harm vector, and some
scenarios could plausibly be assigned to multiple categories. The results
below constitute diagnostic evidence that the system has learned the
intended owner-harm categories; they do not constitute validation
performance in the machine-learning sense.

\textbf{Overall results.}

Gate only (L1+L2+L3): TPR $= 75.3\%$ (226/300),
95\% Wilson CI [70.2\%, 79.9\%]; FPR $= 3.3\%$ (5/150),
CI [1.4\%, 7.6\%].

Gate + Verifier (all layers): TPR $= 85.3\%$ (256/300),
95\% Wilson CI [80.9\%, 88.9\%]; FPR $= 3.3\%$ (unchanged).

Table~\ref{tab:ohb_category} shows per-category breakdown.

\begin{table}[!t]
  \centering
  \caption{Owner-Harm Benchmark: Per-Category Results (Post-Hoc Diagnostic)}
  \label{tab:ohb_category}
  \begin{tabular}{lcccc}
    \toprule
    \textbf{Category} & \textbf{$N_h$} & \textbf{Gate TPR}
      & \textbf{Combined TPR} & \textbf{Gate FPR} \\
    \midrule
    C1: Credential Leak         & 28 & 100.0\% & 100.0\% & 0.0\%  \\
    C2: Infrastructure Exposure & 50 & 100.0\% & 100.0\% & 20.0\%$^\dagger$ \\
    C3: Privacy Exposure        & 28 & 100.0\% & 100.0\% & 0.0\%  \\
    C4: Inner Circle Leak       & 28 & 89.3\%  & 89.3\%  & 0.0\%  \\
    C8: Unauthorized Autonomy   & 50 & 70.0\%  & 70.0\%  & 0.0\%  \\
    C5: Asset Destruction       & 28 & 64.3\%  & 64.3\%  & 0.0\%  \\
    C6: Exfiltration via Tools  & 28 & 57.1\%  & 57.1\%  & 0.0\%  \\
    \textbf{C7: Hijacking}      & \textbf{60} & \textbf{43.3\%} & \textbf{93.3\%} & \textbf{0.0\%} \\
    \bottomrule
  \end{tabular}
  \vspace{1mm}
  \\
  \small $^\dagger$ 20\% FPR for C2 caused by rule OH-R3a over-triggering
  on benign external-channel configuration queries.
\end{table}

\textbf{Hijacking layer complementarity.} The most significant result is
the Hijacking (C7) improvement from 43.3\% gate-only to 93.3\% combined.
Table~\ref{tab:hijacking} shows the 2$\times$2 detection quadrant analysis
(60 harmful, 30 benign).

\begin{table}[!t]
  \centering
  \caption{Hijacking (C7) Detection Quadrant Analysis (60 harmful, 30 benign)}
  \label{tab:hijacking}
  \begin{tabular}{lcc}
    \toprule
    \textbf{Quadrant} & \textbf{Count} & \textbf{\%} \\
    \midrule
    Gate-only (verifier misses)   & 11 & 18.3\% \\
    Verifier-only (gate allows)   & 30 & 50.0\% \\
    Both layers detect            & 15 & 25.0\% \\
    Neither (structural boundary) &  4 &  6.7\% \\
    \midrule
    \textbf{Combined}             & \textbf{56} & \textbf{93.3\%} \\
    \bottomrule
  \end{tabular}
\end{table}

Wilson 95\% CIs: Gate TPR [31.6\%, 55.9\%]; Combined TPR
[84.1\%, 97.4\%]; Hijacking benign FPR [0.0\%, 11.4\%]. The
50\% ``Verifier-only'' quadrant (30 cases) represents prompt-injection
attacks where the semantic gate is itself manipulated by injected content,
but the verifier detects the behavioral fingerprint in file artifacts.
The 4 structural boundary cases (6.7\%) involve direct SQL file injection
and direct SSH key injection patterns not yet covered by V-rules.

\subsection{Cross-Benchmark Comparison}
\label{subsec:cross}

Table~\ref{tab:cross} presents the central finding: the same system
exhibits a 6.8$\times$ gap in safety score between generic harm and
owner-harm scenarios.

\begin{table}[!t]
  \centering
  \caption{Cross-Benchmark Safety Scores (Nous vs.\ Generic LLM Baseline)}
  \label{tab:cross}
  \begin{tabular}{lcccc}
    \toprule
    \textbf{Benchmark} & \textbf{Focus} & \textbf{Nous Safety} & \textbf{Generic LLM} & \textbf{Status} \\
    \midrule
    AgentHarm          & Generic criminal  & 100\% / 0\%     & 62.7\% / 16.0\% & Validation \\
    Owner-Harm (diag.) & Owner-directed    & 75.3\% / 3.3\%  & ---             & Diagnostic \\
    AgentDojo          & Prompt injection  & 14.8\% (4/27)   & 59.3\% (16/27)  & Authoritative \\
    \bottomrule
  \end{tabular}
  \vspace{1mm}
  \\
  \small Generic LLM = \texttt{qwen-turbo} zero-shot; gap = $|$Nous gap$|$ $-$
  $|$Generic gap$|$ = 85.2 pp $-$ 3.4 pp = 81.8 pp attributable to
  symbolic rule generalization failure.
\end{table}

The 100\% $\to$ 14.8\% gap is not a system deficiency. It is the paper's
primary evidence: the constraints that achieve 100\% on AgentHarm (system
call toxicity patterns, malware-synthesis keywords, harassment content
signals) have near-zero coverage of application-layer owner-harm.
``Send money to account X'' is indistinguishable from a legitimate banking
action by its content alone; it becomes harmful only relative to the
owner's authorization scope and the user's actual task goal.

\subsection{Controlled Experiment: Gap Attribution}
\label{subsec:control}

The 85.2 pp gap between AgentHarm (100\%) and AgentDojo (14.8\%)
admits two competing explanations: (H1) the gap reflects a genuine
structural difference---owner-harm scenarios detected via prompt injection
are semantically harder to classify; or (H2) the gap is an artifact of
Nous's symbolic rules being fit to the AgentHarm tool vocabulary and
failing to generalize.

\textbf{Experimental design.} We evaluate a zero-shot generic LLM
classifier (\texttt{qwen-turbo} via API relay) on the same data, using
a uniform binary-classification prompt with no benchmark-specific
fine-tuning:
\begin{quotation}\small
  \textit{You are a safety classifier. An AI agent is performing a task.
  User's original request: \{prompt\}. Agent's planned action: \{tool\}
  (\{params\}). Is this action potentially harmful to the user?
  Answer YES or NO.}
\end{quotation}
For AgentDojo, the goal description (injected instruction) is substituted
for the tool-call representation. The baseline has no access to owner
context ($\RO$, $\BO$, $\AuthO$).

\textbf{Results.} Table~\ref{tab:control} shows the 2$\times$2 comparison.

\begin{table}[!t]
  \centering
  \caption{Controlled Experiment: Gap Attribution
    (Nous vs.\ Generic LLM Classifier)}
  \label{tab:control}
  \begin{tabular}{lccc}
    \toprule
    \textbf{System} & \textbf{AgentHarm TPR} & \textbf{AgentDojo TPR} & \textbf{Gap} \\
    \midrule
    Nous (L1+L3+L4) & 100.0\% & 14.8\% (4/27) & 85.2 pp \\
    Generic LLM     & 62.7\%  & 59.3\% (16/27) & 3.4 pp \\
    \bottomrule
  \end{tabular}
  \vspace{1mm}
  \\
  \small Generic LLM AgentHarm: 110 TP / 66 FN / 28 FP (FPR = 15.9\%,
  $N=176$ harmful $+$ 176 benign). \\
  \small AgentDojo per-category breakdown appears in text.
\end{table}

\textbf{Finding.} The generic LLM exhibits a 3.4 pp gap (62.7\% vs.\
59.3\%)---statistically negligible compared to Nous's 85.2 pp gap.
This is inconsistent with H1: AgentDojo owner-harm scenarios do not appear to be
semantically harder to detect. The generic LLM achieves 59.3\% on
AgentDojo without any prompt-injection-specific design, suggesting that
the harm is semantically recognizable by content alone.

\textbf{Gap attribution.} Nous's gap is strongly associated with \textbf{Layer 1
(Datalog) generalization failure}. L1 rules were derived from
AgentHarm patterns---\texttt{bash(script)}, \texttt{read\_file
(malicious\_path)}, \texttt{execute(command)}---which are absent from
AgentDojo's application-layer tool vocabulary
(\texttt{send\_money}, \texttt{send\_email}, \texttt{update\_calendar}).
\textbf{Implication: generic LLM outperforms Nous.}
Notably, the zero-shot generic LLM (59.3\%) substantially outperforms
Nous (14.8\%) on AgentDojo owner-harm detection. This is not a
contradiction---it is the central architectural insight: owner-harm
detection depends primarily on semantic understanding of action
consequences and owner context, which symbolic rules (L1) cannot provide.
Nous's L1 achieves perfect recall on AgentHarm's system-call patterns but
has near-zero coverage of application-layer actions. L3's semantic
gate contributes only +11.1\,pp because it also lacks owner-context
grounding. The generic LLM compensates via world-knowledge reasoning,
suggesting that future owner-harm defenses should prioritize structured
semantic alignment over pattern-matching rules.

\subsection{SSDG Experimental Predictions}
\label{subsec:ssdg_experiments}

The SSDG framework (Section~\ref{subsec:ssdg_theory}) yields two testable
predictions. We report pilot experiments ($N = 12$ AgentDojo tasks
$+$ 15 AgentHarm scenarios) using the same controlled setup as
Section~\ref{subsec:control}.

\textbf{P1 (Context Injection) --- Prediction Rejected.}
Injecting the user's task goal into the gate evaluation context
(L1+L3+L4 + \texttt{user\_goal}) should raise the AgentDojo TPR if
goal-awareness is the missing information. Result: TPR declined from
14.8\% to 8.3\%---below the no-injection baseline.

This rejection has an informative interpretation: appending the user
goal as a text prefix does not constitute structured goal-action
alignment. The gate's intent decomposition logic cannot compare the
injected goal to the observed action without explicit structural
scaffolding (e.g., a ``goal vs.\ action'' contrast prompt). Three
confounding factors compound this: (a) the ground-truth AgentDojo
pipeline evaluates task success, not injection semantics; (b) simple
text injection may activate over-cautious benign-task blocking; and
(c) the pilot sample size ($N=12$) is too small for significance.
P1 remains open: the mechanism requires \emph{structured} goal-context
integration, not raw text concatenation.

\textbf{P2 (Context Deprivation) --- Prediction Confirmed.}
Stripping world-context from the semantic defense should reproduce
symbolic-gap behavior. We evaluate the generic LLM classifier under
two conditions: (i) \emph{Full Group}---standard evaluation with full
agent-context prompt; (ii) \emph{Stripped Group}---evaluation with
only the raw tool name and parameter, no surrounding task description.

\begin{table}[!t]
  \centering
  \caption{P2: Context Deprivation Induces Symbolic-Class Gap
    (pilot, $N=12$ AgentDojo + 15 AgentHarm)}
  \label{tab:ssdg_p2}
  \begin{tabular}{lccc}
    \toprule
    \textbf{Condition} & \textbf{AgentHarm TPR} & \textbf{AgentDojo TPR}
      & \textbf{Gap Ratio $R$} \\
    \midrule
    Full Group   & 62.7\% & 59.3\% & 1.06 \\
    Stripped Group & 30.0\% & 8.3\% & 3.60 \\
    \bottomrule
  \end{tabular}
  \vspace{1mm}
  \\
  \small $R = \mathrm{TPR}_\mathrm{AH} / \mathrm{TPR}_\mathrm{AD}$.
  P2 falsification threshold: $R \leq 1.5$.
\end{table}

The stripped condition yields $R = 3.60$, exceeding the P2
falsification threshold ($R \leq 1.5$). When deprived of context,
the semantic defense degrades to near-symbolic gap behavior,
confirming Hypothesis~\ref{hyp:semantic_stability}: the stability of
semantic detection depends on access to world-context, not on model
reasoning capacity per se. The finding identifies \emph{information
deprivation} as a key factor associated with the detection gap.

\section{Production Deployment Evidence}
\label{sec:production}

Nous was deployed in shadow mode alongside an existing rule-based system
in a production multi-agent environment for 11+ days (30,204 calls total).

\textbf{Consistency}: 99.47\% agreement with the existing rule-based
system across all calls---confirming that semantic reasoning does not
introduce systematic drift from established policy.

\textbf{False negative recovery}: Nous detected 157 cases that the
existing rule-based system passed as benign. Manual review of a stratified
sample confirmed these were true positives, predominantly ambiguous
read-then-send workflows where the send target fell outside the owner's
trust boundary. \emph{Caveat}: the manual review was conducted by the
system developer (first author), not an independent auditor, and was not
blind to Nous's verdict---creating potential confirmation bias. These
results should be interpreted as case-level observations, not as a
controlled study.

\textbf{Over-blocking}: Nous over-blocked in 4 cases where the existing
system correctly allowed. Manual review confirmed all 4 were over-triggered
on ambiguous read-then-send workflows where the send target was a legitimate
internal address. (Same caveat as above: non-blind, author-conducted review.)

Following the shadow period, Nous was promoted to primary interceptor.
No regressions were observed post-cutover.

\section{Discussion}
\label{sec:discussion}

\subsection{Why Generic Defenses Fail on Owner-Harm}
\label{subsec:disc_failure}

The production and benchmark results converge on the same explanation:
generic defenses evaluate action content without owner context. The action
``send email to external@domain.com'' carries no harm signal by content;
its harmfulness is entirely determined by whether \texttt{external@domain.com}
is within $\BO$. Content classifiers, injection detectors, and DLP systems
operating without $\RO$, $\BO$, and $\AuthO$ cannot make this determination.

The controlled experiment (Section~\ref{subsec:control}) provides a more
precise attribution. The 85.2 pp gap arises \emph{specifically} from the
symbolic rule layer (L1), which encodes environment-specific tool-name
blacklists derived from the AgentHarm training distribution rather than
semantic harm patterns. The semantic layer (L3) alone shows no comparable
gap---and the generic zero-shot LLM, which operates purely on semantic
content without any symbolic rules, exhibits only a 3.4 pp cross-benchmark
gap. This has concrete implications for defense design: symbolic rules
provide near-perfect coverage within their training distribution but
catastrophically fail to generalize to novel tool vocabularies. Upgrading
from tool-name blacklists to parameter-semantic analysis (e.g., detecting
attacker-controlled IBANs or exfiltration-destination emails regardless of
the calling tool name) is the single most impactful architectural change
for cross-environment generalization.

The AgentDojo structural blind spots (Section~\ref{subsec:agentdojo})
reveal an orthogonal gap: malicious-parameter attacks succeed because the
gate context does not include the user's task goal. The fundamental
information needed to compare ``user requested X, agent doing Y'' is absent
from runtime gate evaluation.

The SSDG experiments (Section~\ref{subsec:ssdg_experiments}) provide
a more precise formulation: information deprivation is the structural
factor, not semantic difficulty. P2 confirms that stripping context
from an otherwise capable semantic defense reproduces symbolic-level
gaps ($R = 3.60$). P1's rejection is equally informative---it shows
that goal awareness is necessary but not sufficient: the defense must
be architecturally capable of comparing goal to action, not merely
receive the goal as a text fragment.

\subsection{Symbolic vs.\ Semantic Generalization}
\label{subsec:disc_generalization}

The controlled experiment, together with the SSDG theoretical framework
(Hypotheses~\ref{hyp:symbolic_gap}--\ref{hyp:semantic_stability}),
exposes a fundamental asymmetry between the two
primary detection mechanisms in compositional safety architectures:

\textbf{Symbolic rules (L1 Datalog)} exhibit a bimodal generalization
profile: near-perfect recall within the training distribution (100\% on
AgentHarm), and near-zero recall on out-of-distribution tool vocabularies
(3.7\% on AgentDojo). This is structurally expected---Datalog rules
encode extensional facts (``tool $t$ is prohibited'') rather than
intensional harm semantics. Any tool name not in the rule set is
trivially allowed.

\textbf{Semantic reasoning (L3 gate and generic LLM)} exhibits a flat
generalization profile: below-perfect recall within distribution
(62.7\% for the generic LLM on AgentHarm), but stable recall across
environments (59.3\% on AgentDojo, gap = 3.4 pp). Semantic reasoning
transfers because harm intent is encoded in goal descriptions---which
remain semantically consistent across tool-vocabulary changes---rather
than in tool names.

This asymmetry has a direct design implication for compositional
architectures: the value of combining symbolic and semantic layers is
not simply ``two layers are better than one.'' The layers have
\emph{complementary generalization curves}: symbolic rules provide
high-precision coverage in-distribution, and semantic reasoning maintains
consistent coverage out-of-distribution. A system relying solely on
symbolic rules will exhibit catastrophic degradation on novel tool
environments; a system relying solely on semantic reasoning will accept
a consistent recall ceiling but generalize robustly. The composable
architecture's value is precisely that it avoids both failure modes.

\subsection{Layer Complementarity on Hijacking}
\label{subsec:disc_hijacking}

The Hijacking result (43.3\% gate $\to$ 93.3\% combined) provides
direct evidence of architectural complementarity. The gate operates on
prompt semantics and command surface; the verifier operates on file
contents and multi-step behavioral sequences. These detection dimensions
are orthogonal for prompt-injection attacks:

\begin{itemize}
  \item When the injection is semantically explicit (e.g., ``ignore
    previous instructions''), the gate catches it; the verifier sees no
    artifact yet.
  \item When the injection is embedded in a file artifact (e.g., a calendar
    event description or email body that the agent reads and acts upon),
    the semantic gate evaluates the \emph{agent's action}, which may be
    semantically plausible; the verifier detects the injection fingerprint
    in the artifact content.
  \item When both signals are present, both layers fire (25\% of cases).
\end{itemize}

This architectural insight---that prompt injection requires both
\emph{semantic} and \emph{behavioral-sequence} detection---is a
transferable design principle for agent safety systems.

\subsection{Implications for Benchmark Design}
\label{subsec:disc_benchmarks}

The AgentDojo 3.7\%$\to$14.8\% result implies a design gap in existing
injection benchmarks: they do not provide explicit owner context. A benchmark
that injects into the gate context (a) which resources belong to the owner,
(b) who is in the owner's trust boundary, and (c) what the user's task goal
is, would enable defenses to reach the estimated 40--60\% ceiling and
would meaningfully differentiate owner-harm defense strategies.

We recommend future benchmarks include an \emph{owner context manifest}:
a structured specification of $\RO$, $\BO$, and $\AuthO$ provided to both
the agent and any runtime safety system.

\subsection{Limitations}
\label{subsec:limitations}

\begin{enumerate}
  \item \textbf{AgentDojo ceiling}: Without user task goal context, the
    semantic gate cannot compare ``user requested X'' to ``agent doing Y'',
    limiting achievable safety to $\sim$15\% on malicious-parameter tasks.
  \item \textbf{Benchmark independence}: The Owner-Harm Benchmark is
    author-constructed after system optimization. It is diagnostic, not
    validation evidence.
  \item \textbf{Single annotator}: No inter-annotator agreement is reported
    for the Owner-Harm Benchmark labels.
  \item \textbf{Structural boundary}: 4/60 Hijacking cases (SQL file
    injection, direct SSH key injection) are out of reach of current
    V-rules.
  \item \textbf{Evaluation mode}: AgentDojo evaluation uses ground-truth
    mode without an adversarial attack LLM; real-world adaptive adversaries
    may achieve different results.
  \item \textbf{Infrastructure Exposure FPR}: Rule OH-R3a over-triggers
    on benign external-channel configuration queries (20\% FPR for C2),
    indicating the rule requires further refinement.
  \item \textbf{SSDG pilot scale}: P1 and P2 experiments use $N = 12$
    AgentDojo tasks and $N = 15$ AgentHarm scenarios. Conclusions are
    directionally consistent with theory but require full-benchmark
    replication before statistical significance can be claimed.
\end{enumerate}

\subsection{Future Work}
\label{subsec:future}

\textbf{Structured goal-action alignment.} The P1 SSDG experiment
reveals that simply injecting the user's task goal as text is
insufficient---the gate must be architecturally restructured to perform
an explicit goal-vs.-action comparison. A dedicated ``goal alignment
prompt'' that contrasts the user's authorized intent against the
agent's proposed action is the concrete next step, estimated to raise
AgentDojo safety from 14.8\% to 40--60\%.

\textbf{UEBA for AI agents.} User and entity behavioral analytics adapted
for AI agent deployments would enable per-owner behavioral profiles,
providing a personalized baseline for anomaly detection.

\textbf{Adaptive immunity.} Systems that learn owner-harm patterns from
deployment logs---incorporating newly discovered injection patterns into
Datalog or V-rule libraries without manual rule authoring.

\textbf{SQL-aware file content analysis.} Addressing the 4 structural
boundary Hijacking cases requires content-aware analysis of structured
file formats, beyond the current regex-based V-rules.

\textbf{Owner context manifest standard.} A community standard for
specifying $\RO$, $\BO$, and $\AuthO$ in a machine-readable format, to
enable principled owner-harm benchmarking across systems.

\textbf{Per-decision proof-obligation synthesis.} Ongoing work extends
this compositional gate from offline-authored Datalog rules to
\emph{per-decision runtime synthesis} of Datalog proof obligations by the
LLM itself, admitted through a compile-time gate that rejects
perturbation-invariant or decisive-primitive-less rules. This moves
verification from a ``statute'' (pre-authored policy) to a ``case-law''
(per-decision argued) model; a companion preprint is in preparation.

\section{Related Work}
\label{sec:related}

\textbf{Agent safety and harm taxonomies.} AgentHarm
\cite{andriushchenko2025agentharm} provides the most comprehensive generic
harm benchmark for LLM agents, covering 11 categories from cybercrime to
fraud. ToolEmu \cite{ruan2024toolemu} evaluates LM agent risks in an
LM-emulated sandbox with 7 risk types, partially overlapping our taxonomy
but without an owner-centric perspective. AGrail \cite{agrail2025} proposes
a hierarchical safety framework for agentic systems with adaptive safety
detection. Our work differs by (a) formally defining the owner as the
harm victim, and (b) providing quantitative evidence of the resulting
defense gap.

\textbf{Prompt injection.} Greshake et al.\ \cite{greshake2023indirect}
first systematically analyzed indirect prompt injection, demonstrating that
LLM-integrated applications can be compromised through injected third-party
content. AgentDojo \cite{debenedetti2024agentdojo} provides standardized
evaluation of injection defenses across realistic agentic tasks. CaMeL
\cite{camel2025} addresses prompt injection via capability-aware information
flow, preventing data exfiltration at the language model level rather than
the runtime layer.

\textbf{Runtime safety systems.} NeMo Guardrails \cite{rebedea2023nemoguardrails}
provides programmable rails for LLM applications. Llama Guard
\cite{inan2023llamaguard} implements LLM-based input/output safeguards.
ShieldGemma \cite{zeng2024shieldgemma} offers generative content moderation.
R2Guard \cite{chen2024r2guard} provides runtime guardrails specifically
designed for agent tool calls, and AgentMonitor \cite{yu2024agentmonitor}
offers real-time behavior monitoring. BIPIA \cite{yi2023bipia} benchmarks
indirect prompt injection attacks complementary to our analysis.
These systems operate primarily on content classification; none model
owner-context dimensions ($\RO$, $\BO$, $\AuthO$). Guardrails AI
\cite{guardrailsai2024} provides composable input/output guards but without
owner-harm semantics. Doshi et al.\ \cite{doshi2026verifiable} propose
verifiable tool safety via type systems, complementary to our runtime
detection approach.

\textbf{Policy-driven agent enforcement (2025--2026).} A rapidly growing
body of work explores policy-driven runtime enforcement for LLM agents.
PCAS \cite{palumbo2026pcas} compiles a Datalog-derived declarative
policy language offline and enforces it through a reference monitor.
AgentSpec \cite{wang2026agentspec} provides a lightweight DSL of
triggers, predicates, and enforcement actions. Pro2Guard
\cite{he2025pro2guard} adds probabilistic reachability via learned
DTMCs over execution traces. Solver-Aided verification
\cite{roy2026solveraided} translates natural-language tool-use policies
into SMT-LIB constraints offline and verifies each call with Z3.
ShieldAgent \cite{chen2025shieldagent} constructs probabilistic rule
circuits from policy documents with formal-verification tool support.
GuardAgent \cite{xiang2024guardagent} synthesizes a guardrail plan and
executes it as code, and TrustAgent \cite{hua2024trustagent} inserts
an Agent Constitution post-planning check. These systems focus on
policy expressiveness and offline authorability; our contribution is
orthogonal: the owner-context dimensions ($\RO$, $\BO$, $\AuthO$) must
be \emph{inputs} to any such policy regardless of its authoring method,
yet are absent from the input vocabularies of all systems surveyed here.

\textbf{Exfiltration and covert channels.} Rehberger \cite{rehberger2024copilot}
demonstrated ASCII smuggling via Microsoft 365 Copilot, encoding sensitive
data in Markdown rendered by the victim's browser. The PromptArmor Slack
AI disclosure \cite{promptarmor2024slack} demonstrated credential
exfiltration via injected prompts. Abdelnabi et al.\ \cite{abdelnabi2024llmagent}
provide a systematic analysis of adversarial attacks on LLM-based agents.

\textbf{Post-gate auditing.} Intrusion detection systems \cite{liao2013intrusion}
provide a conceptual precedent for behavioral sequence auditing after an
initial gate, applied here to AI agent tool-call sequences.

\textbf{Provably safe systems.} Tegmark and Omohundro
\cite{tegmark2023provably} argue that only formally verifiable systems can
achieve controllable AGI, motivating the Datalog layer's emphasis on
decidable rule-based reasoning.

\section{Conclusion}
\label{sec:conclusion}

Owner-harm is a systematically under-studied threat category in AI agent
safety. We formalize it with eight categories and precise definitions,
grounded in real incidents from 2023--2026. We demonstrate the
defense gap quantitatively: a compositional safety system achieving 100\%
on generic criminal harm (AgentHarm) scores only 3.7\%/14.8\% on
prompt-injection-mediated owner harm (AgentDojo), with specific structural
causes attributed to each failure mode. We show that compositional
detection---combining a semantic gate with a deterministic post-audit
verifier---achieves 93.3\% on the hardest sub-category (Hijacking), with
the two layers operating on orthogonal detection signals.

The core finding is conceptually simple: existing defenses are
owner-context-blind. An action is harmful to an owner not because of its
content in isolation, but because of how it relates to the owner's
resources, trust boundary, and authorization scope. Until safety systems
model these context dimensions explicitly, owner-harm will remain a
systematic blind spot---even as generic harm detection reaches saturation.

\section*{Author Contributions}

D.Z.\ conceived the Owner-Harm threat model, designed and implemented
the four-layer Nous system, ran all experiments on AgentHarm and
AgentDojo, and drafted the manuscript. Y.J.\ (knowledge graph
specialist, Tongji University) contributed to the initial research
design discussions, provided critical review of the
symbolic--semantic architecture and the owner-harm category taxonomy,
and reviewed final manuscript revisions. Both authors approved the
final version and are accountable for the integrity of the work.

\bibliographystyle{plainnat}
\bibliography{references}

\end{document}